\def\o{\over}
\def\A{\rightarrow}

\def\r{\gamma}
\def\d{\delta}
\def\a{\alpha}
\def\b{\beta}
\def\n{\nu}
\def\e{\epsilon}
\def\p{\pi}
\def\th{\theta}
\def\om{\omega}
\def\vp{{\varphi}}

\def\Im{{\rm Im}}

\def\t{\tilde}

\documentstyle [12pt]{article}
\begin{document}
\baselineskip=25pt
\setcounter{page}{1}
\thispagestyle{empty}
\topskip 1  cm
\topskip 0.0  cm
\begin{flushright}
\begin{tabular}{c c}
& {\normalsize   UWThPh-1994-53}\\
& \today
\end{tabular}
\end{flushright}
\vspace{1 cm}
\centerline{\Large\bf See-saw Enhancement of  Neutrino Mixing}
\centerline{\Large\bf due to the Right-handed Phases }
\vskip 2 cm
\centerline{{\bf Morimitsu TANIMOTO}
  \footnote{Permanent address:Science Education Laboratory, Ehime University,
79
   0 Matsuyama, JAPAN}}
\vskip 1 cm
 \centerline{ \it{Institut f\"ur Theoretische Physik,
               Universit\"at Wien}}
\centerline{ \it Boltzmanngasse 5, A-1090 Wien, AUSTRIA}

\vskip 2.5 cm
\centerline{\bf ABSTRACT}
\vskip 0.5 cm
 We study the see-saw enhancement mechanism
 in presence of the right-handed phases of the Dirac neutrino
 mass matrix and the Majorana mass matrix.  The enhancement condition given by
S
   mirnov  is modified.  We point out that the see-saw enhancement could be
obt
   ained due to the right-handed phases
even if the Majorana matrix is
proportional to the unit matrix.  We show a realistic Dirac mass matrix which
ca
   uses the see-saw enhancement.
\newpage
\topskip 1 cm
 The see-saw mechanism of the neutrino mass generation gives a very natural and
   elegant understanding for the smallness of neutrino masses[1]. The recent
obs
   erved solar neutrino deficit[2] and   muon neutrino deficit in the
atomospher
   ic neutrino flux[3] s
timulate the systematic study of the neutrino mixings[4].
In the standpoint of the quark-lepton unification in most GUT groups, the Dirac
   mass matrix of neutrinos is similar to the one of quarks, therefore, the
neut
   rino mixings turn out to be typically of the same order of magnitude as the
q
   uark mixings.  Howeve
r, some authors pointed out that the large neutrino mixing between the
different
    generations could be obtained in the see-saw mechanism as a consequence of
c
   ertain structure of the right-handed Majorana mass matrix[5,6,7]. That is
the
    so called see-saw en
hancement[7] of the neutrino mixing due to the cooperation between the Dirac
and
    Majorana mass matrices, which needs a higher representation
of the Higgs fields such as {\bf 126} in SO(10).\par
   In this paper, we modify the enhancement conditions[7], which were given by
S
   mirnov in presence of the right-handed phases of the Dirac mass matrix and
th
   e Majorana mass marix. So,
 we point out that the see-saw enhancement could be  obtained even if the
Majora
   na matrix is
proportional to the unit matrix, i.e., there is no hierarchy in the Majorana
mas
   s matrix.  This enhancement is caused by the right-handed phases, which
never
appear in the case of the quark mixing. The large enhancement is obtained  for
s
   ome textures of the Dirac mass matrix.
\par
 In order to see the role of the right-handed phases of the Dirac mass matrix
cl
   early,
 we work in the nearest neighbour interaction(NNI) basis[8], which
is one specific weak basis and does not imply any loss of generality for the
Dir
   ac  mass matrix. In this basis, the Dirac
mass matrix with three generations is written as follows:

  \begin{equation}
 D_i = \left( \matrix{0 & A_i e^{i\a} & 0\cr A'_i e^{i\a'} & 0 &
   B_i e^{i\b}\cr 0 & B'_i e^{i\b'} & C_i e^{i\r}\cr} \right )  \ ,
\end{equation}
\noindent
 where $i$ denotes quarks or leptons, and
  $A_i$, $B_i$ and $C_i$  are positive real numbers.  Using the diagonal phase
m
   atrices $P_L$ and $P_R$, the Dirac mass matrix turns to be real one $R_i$
suc
   h as
 \begin{equation}
  D_i = P_L^* \left( \matrix{0 & A_i  & 0\cr A'_i  & 0 &  B_i \cr
       0 & B'_i  & C_i \cr} \right ) P_R=P_L^* R_i P_R   \ ,
\end{equation}
\noindent where
\begin{equation}
  P_L = \left( \matrix{ e^{-i(\a-\b')} & 0 & 0 \cr
      0 & e^{-i(\b-\r)}  & 0  \cr   0 & 0 & 1 \cr} \right )  \ ,
\qquad
  P_R = \left( \matrix{ e^{i(\a'-\b+\r)} & 0 & 0 \cr
      0 & e^{i\b'} & 0 \cr 0 & 0 & e^{i\r} \cr} \right )  \ .
\end{equation}
\noindent
These phase matrices are removed by re-defining the left-handed and
right-handed
    fermion fields.  Then, the phase matrix $P_L$  appears  in the expression
of
    the quark
  mixing matrix.   The right-handed phase matrix $P_R$ never appear in the
quark
    mixings
since the charged current is the left-handed one. \par
However, neutrino mixings with the see-saw mechanism
 are affected by the right-handed phase matrix $P_R$
because the  Majorana mass matrix is the right-handed one.
This point is an important difference from the case of the quark mixings.  In
th
   e basis of the real Dirac mass matrix,
 the see-saw mass matrix is  written as follows:
\begin{equation}
    \left( \matrix{0 & R_N\cr R_N^T  & P_R^* M_R P_R^*
                         \cr} \right )  \ ,
\end{equation}
\noindent where $M_R$ is a right-handed Majorana matrix.
  It should be noticed that  $M_R$ is multiplied by
 $P_R^*$ in both sides because $M_R$ is  Majorana mass.
Since the matrix $M_R$ may be a complex off-diagonal matrix,
 it is generally written
as $\t P_R^* \t U_R M_R^{diag}\t U_R^T \t P_R^*$, where
  $\t P_R^*$, $\t U_R$
 and $M_R^{diag}$ are the phase matrix, the orthogonal one and
 the diagonal one, respectively.
 If we move to the diagonal basis of the Dirac mass matrix
 by a bi-orthogonal transformation such as
 $U_L^T R_N U_R\A R_N^{diag}$,
 the see-saw mass matrix turns to be
\begin{equation}
   \left( \matrix{0 & R_N^{diag} \cr R_N^{diag}   &
 U_R^T P_R^* \t P_R^* \t U_R M_R^{diag}\t U_R^T\t P_R^* P^*_R U_R \cr} \right )
    \ .
\end{equation}
\noindent
 Then, the light neutrino mass matrix $m_\n$ is given
 as follows:
\begin{eqnarray}
 m_{\n} &=& R_N^{diag}(U_R^T P_R^* \t P_R^* \t U_R M_R^{diag}
   \t U_R^T \t P_R^*P^*_R U_R)^{-1} R_N^{diag} \nonumber \\
 &=& R_N^{diag} U_R^T Q_R \t U_R
     (M_R^{diag})^{-1} \t U_R^T Q_R U_R R_N^{diag}  \ ,
\end{eqnarray}
\noindent
where $P_R \t P_R$ is replaced with a single phase matrix $Q_R$.
This mass matrix is a diagonal one
   only if $U_R^T  Q_R \t U_R
      (M_R^{diag})^{-1} \t U_R^T Q_R U_R$
 is proportional to the unit matrix.  In this case, the lepton mixing  matrix
is
    obtained by diagonalizing the charged lepton mass matrix. Therefore, the
mix
   ing is expected to be of the same order of magnitude as the quark mixings.
However, even if $M_R^{diag}$ is proportional to the unit matrix, i.e., the
case
    of the three degenerated right-handed Majorana masses, the mass matrix
$m_{\
   n}$ is not diagonal one unless  the phase matrix
  $Q_R$ is a real matrix except for an over all phase.
Since the phase elements in $Q_R$
 are independent of ones in $P_L$ in eq.(3),
  these parameters are new freedoms
of the lepton mixings in contrast to the case of the
 quark mixings.
Generally, the lepton mixing matrix is given as
 $N_L^T U^T_L P_L C_L$, where $N_L$ is the unitary
matrix to diagonalize  $m_{\n}$ in eq.(6)
 and $C_L$ is the orthogonal
matrix to diagonalize the real charged lepton mass matrix
in the NNI basis.
The CP violating phase could be  in both $P_L$ and $N_L$.
\par
   In order to show the effect of the phase matrix $Q_R$, we
    consider
 two-generation case.
 We parametrize following matrices
in the two generations,
\begin{eqnarray}
U_R &=&  \left( \matrix{\cos\th^D_R & \sin\th^D_R \cr
-\sin\th^D_R &  \cos\th^D_R  \cr} \right ) \ , \qquad
\t U_R =  \left( \matrix{\cos\th^M & \sin\th^M \cr
-\sin\th^M &  \cos\th^M  \cr} \right ) \ , \nonumber \\
Q_R&=& \left( \matrix{1 & 0 \cr 0 & e^{i\phi}  \cr} \right ) \ , \qquad
M_R^{diag}= \left( \matrix{M_1 & 0 \cr 0 & M_2  \cr} \right ) \ ,   \qquad
R_N^{diag}=\left( \matrix{m_1 & 0 \cr 0 & m_2 \cr} \right ) \ .
\end{eqnarray}
\noindent
 The light neutrino mass matrix $m_\nu$ of eq.(6) is
 given by six parameters,  $m_1$, $M_1$, $m_2^2/M_2$, $\th^D_R$, $\th^M$ and
$\p
   hi$.  Since this matrix is very long matrix and somewhat complicated one, we
   show it
 in the basis of $\th^M=0$, in other words, the diagonal basis of the Majorana
m
   ass matrix  $M_R$.
 Defining mass hierarchy parameters such as
 \begin{equation}
\e^D\equiv {m_1\o m_2} \ , \qquad\qquad
\e^M\equiv {M_1\o M_2} \ ,
\end{equation}
 we get
\begin{equation}
m_{\n}=\left ( \matrix{|m_{11}|e^{i \a} & |m_{12}|e^{i \b} \cr
  |m_{12}|e^{i \b} & |m_{22}|e^{i \r}\cr }\right ) \  ,
\end{equation}
\noindent
where
\begin{eqnarray}
 |m_{11}|&=&  {(e^D m_2)^2\o \e^M M_2 }
 [\cos^4 \th^D_R+(e^M)^2\sin^4\th^D_R+2e^M\sin^2\th^D_R\cos^2\th^D_R
\cos 2\phi]^{1\o 2} \ , \nonumber \\
|m_{21}|&=& {e^D m_2^2\o \ 2 e^M M_2}
 [1+(e^M)^2-2e^M\cos 2\phi]^{1\o 2}\sin 2\th^D_R \ , \nonumber \\
|m_{22}|&=&  { m_2^2\o \e^M M_2 }  [\sin^4
\th^D_R+(e^M)^2\cos^4\th^D_R+2e^M\sin
   ^2\th^D_R\cos^2\th^D_R
\cos 2\phi]^{1\o 2} \ , \nonumber\\
\tan\a&=& {\e^M\sin^2\th^D_R\sin 2\phi\o
     \cos^2\th^D_R+\e^M \sin^2\th^D_R\cos 2\phi} \ , \nonumber\\
\tan\b&=& {-\e^M\sin 2\phi\o
     1-\e^M\cos 2\phi} \ , \nonumber \\
\tan\r&=& {\e^M\cos^2\th^D_R\sin 2\phi\o
     \sin^2\th^D_R+\e^M \cos^2\th^D_R\cos 2\phi} \ .
\end{eqnarray}
\noindent
Some of the phases in the matrix of eq.(9) are removed by the phase rotation of
    the light neutrino fileds as follows:
\begin{equation}
m_{\n} \Rightarrow
  \left ( \matrix{e^{-{i\o 2}(\b-\om)} & 0 \cr
  0 & e^{-{i\o 2}(\b+\om)}\cr }\right ) m_{\n}
  \left ( \matrix{e^{-{i\o 2}(\b-\om)} & 0 \cr
  0 & e^{-{i\o 2}(\b+\om)}\cr }\right )=
\left ( \matrix{|m_{11}|e^{i \varphi_1} & |m_{12}|\quad\ \ \cr
  |m_{12}|\quad\ \ & |m_{22}|e^{i \varphi_2}\cr }\right ) \  ,
\end{equation}
\noindent
 where $\varphi_1=\a-\b+\om$, $\varphi_2=\r-\b-\om$,  and
 $\om$ is fixed in order to give the following relation:
\begin{equation}
 \Im\{|m_{11}|e^{i (\a-\b+\om)}\}=\Im\{|m_{22}|e^{i (\r-\b-\om)}\} \ ,
\end{equation}
\noindent
 which gives
\begin{equation}
\tan\om={|m_{22}|\sin(\r-\b)-|m_{11}|\sin(\a-\b)
        \o |m_{22}|\cos(\r-\b)+|m_{11}|\cos(\a-\b)} \ .
\end{equation}
\par
Since the  phase matrix in eq.(11) is absorbed into the charged
 lepton fields finally[9], it does not affect the  observable
 quantities such as $CP$ violation.  Now, we can diagonalize the neutrino
 mass matrix in eq.(11) by using the orthogonal matrix, while
 those mass eigenvalues are complex ones. One phase in the eigenvalues  cannot
b
   e removed away, and then, causes $CP$ violation even in the two generation
ca
   se[10].
 We get the angle $\th_{ss}$ in the orthogonal matrix, which was called as the
s
   ee-saw  angle by Smirnov[7], as follows:
 \begin{equation}
\tan 2\th_{ss} = -2\tan(-\th^D_R){ \e^D[(1-\e^M)^2
   +2\e^M(1-\cos 2\phi)]^{1\o 2}\o
  [(\tan^2(-\th^D_R)+\e^M)^2-
  2\e^M\tan^2(-\th^D_R)(1-\cos 2\phi)]^{1\o 2}\cos\varphi_2-\d},
\end{equation}
\noindent where $\d$ is a small term of second order in the Dirac mass
hierarchy
   :
 \begin{equation}
  \d=(\e^D)^2[(\e^M\tan^2(-\th^D_R)+1)^2-
  2\e^M\tan^2(-\th^D_R)(1-\cos 2\phi)]^{1\o 2}\cos\vp_1 \ .
\end{equation}
In eqs.(14) and (15),  the minus sign in front of $\th_R^D$ is taken  in order
t
   o compare with  eq.(5) in ref.[7].
The mixing matrix of the leptonic charged currents can be written in the
form[9]
\begin{equation}
 \left [\matrix{\cos(\th_L^D-\th_L^\ell+\th_{ss}) &
\sin(\th_L^D-\th_L^\ell+\th_
   {ss}) \cr
-\sin(\th_L^D-\th_L^\ell+\th_{ss}) & \cos(\th_L^D-\th_L^\ell+\th_{ss}) \cr}
\right ] e^{i \rho \tau_3} \ ,
\end{equation}
\noindent
where $\th_L^D$($\th_L^\ell$) is the angle of the  rotation of the left-handed
n
   eutrino(charged lepton)  components, which diagonarizes the Dirac mass
matrix
    of neutrinos(charged leptons), and  $\rho$ is the $CP$ violating phase,
whic
   h
 could be written in terms of six parameters in eq.(7).
If we choose $\phi=0$, the $CP$ violating phase  $\rho$ disappears, and our
eqs.
   (14) and (15) are reduced to eqs.(5) and (6) in ref.[7] in the basis of
$\th^M=0$. However,  the see-saw enhancement conditions given by Smirnov[7]
sho
   uld be modified because
the phase $\phi$ is not generally non-zero.\par
Let us consider modified conditions of the see-saw enhancement.
According to Smirnov's discussion, there are two possiblities of the see-saw
enh
   ancement: (i) all terms in the denominator of the RHS of the eq.(14) are
very
    small, (ii) there is a strong cancellation in the denominator. For the case
   (i), the effect of th
e $\phi$ is very small because both $\e^M$ and $\tan(-\th_R^D)$
 must be small.  On the other hand,
the phase $\phi$ plays  an important role for the case (ii).
The strong cancellation in the denominator is occured at the minimum value of
th
   e following  term
\begin{eqnarray}
& &(\tan^2(-\th^D_R)+\e^M)^2 -
  2\e^M\tan^2(-\th^D_R)(1-\cos 2\phi) \nonumber \\
& &= (\tan^2(-\th^D_R) + \e^M\cos 2\phi)^2+(\e^M)^2(1-\cos^2 2\phi) \ .
\end{eqnarray}
\noindent
The minimum value is obtained
  such as $(\e^M)^2(1-\cos^2 2\phi)$  at
\begin{equation}
      \tan^2(-\th^D_R)=-\e^M\cos 2\phi\ .
\end{equation}
Then, the see-saw mixing
is given as follows:
\begin{equation}
\tan 2\th_{ss} = {-2|\e^M\cos 2\phi|^{1\o 2}[(1-\e^M)^2
   +2\e^M(1-\cos 2\phi)]^{1\o 2}\o
  \e^M (\e^D)^{-1}(1-\cos^2 2\phi)^{1\o
2}\cos\varphi_2-\e^D[1+(\e^M)^2((\e^M)^2
   -2)\cos^2 2\phi]^{1\o 2}\cos\varphi_1} \ .
\end{equation}
\noindent
 Eqs.(18) and (19) reduce to eqs.(10) and (11) of ref.[7] in the case of
$\phi=0
   $ and $\e^M\ll 1$.\par
 It is remarked that the see-saw angle $\th_{ss}$ is not zero even in the case
o
   f $e^M=1$  as seen in eqs.(14) and (19).
 Especially, if the condition $\tan^2(-\th^D_R)=-\cos 2\phi$
  with $e^M=1$ is satisfied, the see-saw enhancement is obtained as follows:
\begin{equation}
\tan 2\th_{ss} = {-2|\cos 2\phi|^{1\o 2}[2(1-\cos 2\phi)]^{1\o 2}\o
(\e^D)^{-1}
   (1-\cos^2 2\phi)^{1\o 2}\cos\varphi_2-\e^D[1-\cos^2 2\phi]^{1\o
2}\cos\varphi
   _1} \ .
\end{equation}
\noindent Then, the maximal mixing $\th_{ss}=\pi/4$ is obtained
at $\phi=\pi/2$, which gives
 $\cos\vp_1=\cos\vp_2=1$ and $\tan(-\th^D_R)=1$.
This situation is contrast to the one in ref.[7],
 where $\th_{ss}$ is precisely zero in the case of $e^M=1$.\par
 Our formula in eq.(14) is given in the basis of $\th^M=0$
 for simplicity.  Although the general formula is complicated one in the basis
o
   f $\th^M\not= 0$,
   it is noticed that  the case of $\th^M=\th^D_R$ does not lead to
$\th_{ss}=0
   $ due to the phase $\phi$.
 If we take $\phi=\pi$ tentatively, $\tan 2\th_{ss}$ is
 given as the same one
 as eq.(5) in ref.[7] except replacing  $\th^M$  with $-\th^M$.
  Thus, the right-handed phase affects  the see-saw angle $\th_{ss}$
drastically
   .\par
  Modification of the see-saw enhancement condition in ref.[7]
 is minor in the case of $\e^M\ll 1$.
Therefore, we focus  on only the case of $\e^M \simeq 1$
 in the following discussions.
In this case, the see-saw enhancement gives the maximal angle $\th_{ss}=\pi/4$
   at $\phi=\pi/2$ with
 $\tan(-\th^D_R)=1$ as discussed above.
Therefore, we need the Dirac mass matrix to satisfy
the condition $\tan(-\th^D_R)=1$.
  If we take, for example, the Fritzsch texture[11] for the Dirac mass matrix,
$
   \th_R^D$ is small due to the
hierarchical fermion masses such as $\sin\th_R^D\simeq \sqrt{m_1/m_2}$.
 So, the see-saw enhancement is not obtained.
However, we know at least two textures of the Dirac mass matrix
 which lead to the large $\th_R^D$.
One is the democratic mass matrix[12] and the other is the
 non-Hermitian mass matrix, which is the modified Fritzsch texture, proposed by
   Branco and Silva-Marcos[13].
 We can easily present a realistic mass matrix model with
above condition.\par
Let us show an example by using the
non-Hermitian Dirac mass matrix, which was proposed by Branco and
Silva-Marcos[1
   3].
 In NNI basis of eq.(1), the assumptions of
$A'=A$ and $B'=B$ give the Fritzsch mass matrix[11], which
implies correlations between quark masses and mixings.
However, the large top-quark mass is unfavourable to the Fritzsch
 ansatz.
Branco and Silva-Marcos have taken another ansatz[13],
$A'=A$ and $B'=C$, which give the  non-Hermitian mass matrix.
This new mass matrix seems to be consistent with the observed
quark mixing matrix and quark masses[14].
After absorbing phases with $P_L$ and $P_R$ in eq.(3), the  $3\times 3$ real
Dir
   ac mass matrix $R_i$ is written as
 \begin{equation}
  R_i = \left( \matrix{0 & A_i  & 0\cr A_i  & 0 &  B_i \cr
       0 & C_i  & C_i \cr} \right ) \ ,
\end{equation}
\noindent which
is diagonalized by a bi-orthogonal transformation such as
 $U^T_L R_i U_R$.
The orthogonal matrices can be expressed approximately
in terms of quark masses as follows:
\begin{eqnarray}
  U_L &\simeq& \left[ \matrix{1 & -2^{-{1\o 4}}({m_1\o m_2})^{1\o 2}  &
    2^{-{1\o 4}}({m_1m_2\o m_3^2})^{1\o 2}     \cr
    2^{-{1\o 4}}({m_1\o m_2})^{1\o 2}  & 1 & {m_2\o m_3}  \cr
     -2^{3\o 4}({m_1m_2\o m_3^2})^{1\o 2} & -{m_2\o m_3} & 1 \cr} \right ] \ ,
\
   nonumber \\
 U_R &\simeq& \left[ \matrix{1 & -2^{1\o 4}({m_1\o m_2})^{1\o 2}  &
    2^{1\o 4}{m_2\o m_3}({m_1m_2\o m_3^2})^{1\o 2}     \cr
2^{-{1\o 4}}({m_1\o m_2})^{1\o 2} &{1\o\sqrt{2}} & {1\o\sqrt{2}}\cr
     -2^{-{1\o 4}}({m_1\o m_2})^{1\o 2} & -{1\o\sqrt{2}}  & {1\o\sqrt{2}}  \cr}
   \right ] \ ,
\end{eqnarray}
\noindent
where $m_1$, $m_2$ and $m_3$ denote the fermion masses of the
 first, second and third generations, respectively.
As seen in eq.(22), the left-handed orthogonal matrix $U_L$
 has a hierarchical structure, which is consistent with
 the observed quark mixings.  On the other hand,
 the right-handed orthogonal matrix $U_R$
 has a hierarchical structure between the first generation
 and the second one, but a democratic structure between
 the second generation and the third one.
If an unknown right-handed relative phase between the second generation and the
   third one is around $\pi/2$, the neutrino mixing angle between  the second
ge
   neration and the third one becomes large
 due to the see-saw enhancement in the case that
 there is no hierarchy in the Majorana mass matrix.
Thus, the see-saw enhancement due to the right-handed phase
 is possible in the realistic Dirac mass matrix.
\par
  Summary is given as follows:
We studied the see-saw enhancement mechanism
 in presence of  the right-handed phases of the Dirac neutrino
 mass matrix and the Majorana mass matrix.  The enhancement conditions given by
   Smirnov  is modified.  We pointed out that the see-saw enhancement could be
   obtained due to the right-handed phases.
  In that case, the hierarchical structure of the right-handed Majorana mass
mat
   rix is not required to get the see-saw enhancement.   We showed a realistic
e
   xample by using the
non-Hermitian Dirac mass matrix, which was proposed by Branco and Silva-Marcos.
More phenomenological analyses are helpful to confirm this enhancement
mechanism
    in
 other Dirac matrix textures.
\par
\vskip 1 cm
I thank the particle group at Institut f\"ur Theoretisch Physik
in Universit\"at Wien, especially,
 Prof.H. Pietschmann, for kind hospitality.
\newpage
\centerline{\large \bf References}
\vskip 1 cm
\noindent
[1] M. Gell-Mann, P. Ramond and R. Slansky, in {\it Supergravity}, Proceedings
o
   f the \par
Workshop, Stony Brook, New York, 1979, edited by P. van Nieuwenhuizen \par
and D. Freedmann, North-Holland, Amsterdam, 1979, p.315;\par
 T. Yanagida, in {\it Proceedings of the Workshop on the Unified Theories and
Ba
   ryon\par
 Number in the Universe}, Tsukuba, Japan, 1979, edited by O. Sawada and \par
A. Sugamoto, KEK Report No. 79-18,
    Tsukuba, 1979, p.95.\par\noindent
[2] K.S. Hirata et al., Phys. Rev. Lett. {\bf 66}(1991)9;
 Phys. Rev. {\bf D44}(1991)2241;\par
 GALLEX Collaboration, P. Anselmann et al., Phys. Lett.
  {\bf B285}(1992)376, 390;\par
   E. Bellotti, Plenary Talk at 27th International Conference
   on High Energy Physics, \par
   Glasgow, July 1994.\par\noindent
[3] K.S. Hirata et al., Phys. Lett. {\bf B280}(1992)146;\par
 R. Becker-Szendy et al,. Phys. Rev.
      {\bf D46}(1992)3720;\par
 Y. Fukuda et al., Phys. Lett. {\bf B335}(1994)237.\par\noindent
[4] C.H. Albright, Phys. Rev.
     {\bf D43}(1991)R3595, {\bf D45}(1992)R725;\par
 G. Lazaridis and Q. Shafi, Nucl. Phys. {\bf B350}(1991)179;\par
 S.A. Bludman, D.C. Kennedy and P. Langacker, Nucl. Phys.
{\bf B373}(1992)498;\par
 K.S. Babu and Q. Shafi, Phys. Lett. {\bf B294}(1992)235;\par
 S. Dimopoulos, L. Hall and S. Raby, Phys. Rev. Lett.
   {\bf 68}(1992)1984; \par
  Phys. Rev. {\bf D45}(1992)4192;\par
 M. Fukugita, M. Tanimoto and T. Yanagida, Prog. Theor. Phys.
     {\bf 89}(1993)263.\par\noindent
[5] M. Tanimoto, T. Hayashi and M. Matsuda, Z. Phys.
   {\bf C58}(1993)267.\par\noindent
[6] C.H. Albright and S. Nandi,  Fermilab-Pub-94/061-T
      (1994)(unpublished).\par\noindent
[7] A. Yu. Smirnov, Phys. Rev. {\bf D48}(1993)3264;
      IC/93/359(1993)(unpublished).\par\noindent
[8] G.C. Branco, L. Lavoura and F. Mota, Phys. Rev.
    {\bf  D391}(1990)3443.\par\noindent
[9] M. Fukugita and T. Yanagida,
   "{\it Physics of Neutrino}", YITP/K-1050(1993),\par
 to be published in Physics and Astrophysics of Neutrinos,
 Springer-Verlag, Tokyo. \par
\noindent
[10] J. Schechter and J.W.F. Valle,
                  Phys. Rev. {\bf  D22}(1980)2227;\par
 S.M. Bilenky, J. Ho\v sek and S.T. Petcov, Phys. Lett.
   {\bf 94B}(1980)495;\par
 M. Doi et al., Phys. Lett. {\bf 102B}(1981)323.\par
\noindent
[11] H. Fritzsch, Phys. Lett. {\bf 70B}(1997)436;
        Nucl. Phys. {\bf B155}(1979)189.\par\noindent
[12] H. Harari, H. Haut and J. Weyers, Phys. Lett.
   {\bf 78B}(1978)459;\par
P. Kaus and S. Meshkov, Mod. Phys. Lett.{\bf A3}(1988)1251;\par
 Y. Koide, Phys. Rev. {\bf  D28}(1983)252;
                 {\bf  D39}(1989)1391;\par
G.C. Branco, J.I. Silva-Marcos and M.N. Rebelo, Phys. Lett.
     {\bf 237B}(1990)446.\par\noindent
[13]  G.C. Branco and J.I. Silva-Marcos, Phys. Lett.
     {\bf 331B}(1994)390.\par\noindent
[14] Particle Data Group, Phys. Rev. {\bf  D50}(1994)1315.
     \par
\end{document}